\documentclass[aps, preprintnumbers, reprint, eqsecnum, nofootinbib,superscriptaddress]{revtex4-1}
\usepackage{epsf}
\usepackage{epsfig}
\usepackage{graphics}
\usepackage{graphicx}
\usepackage{amssymb}
\usepackage{mathtools}
\usepackage{color}
\usepackage{hyperref}
\usepackage{multirow}
\usepackage{feynmp-auto}
\usepackage[normalem]{ulem}
\usepackage{amsmath, latexsym, amssymb, hyperref, graphicx, slashed, color, multirow}
\definecolor{nicered}{rgb}{0.0,.7,.3}
\definecolor{nicegreen}{rgb}{.1,.5,.1}
\definecolor{darkblue}{rgb}{0,.1,.9}
\definecolor{blue}{rgb}{0,0,0.5}
\definecolor{lightblue}{rgb}{0,0,1}
\definecolor{red}{rgb}{0.5,0,0}
\definecolor{lightred}{rgb}{1,0.5,0}
\definecolor{green}{rgb}{0,0.5,0}
\definecolor{darkgreen}{rgb}{0.0,0.3,0.0}
\definecolor{orange}{rgb}{1,0.4,0}
\definecolor{grey}{rgb}{0.5,0.5,0.5}
\hypersetup{colorlinks, citecolor=nicered ,linkcolor=darkblue, urlcolor=nicered}
\pdfoutput=1
\usepackage{latexsym, amssymb, hyperref, graphicx, slashed, color, multirow}
\usepackage[T1]{fontenc}
\hypersetup{colorlinks, citecolor=nicered ,linkcolor=darkblue, urlcolor=nicered}

\usepackage{graphicx}
\usepackage{amsmath,amssymb}
\usepackage{amsfonts}
\usepackage{adjustbox}
\usepackage{color}
\usepackage{slashed}
\usepackage{float}
\usepackage{feyn}
\usepackage{feyn}
\usepackage{hyperref}

\allowdisplaybreaks

\newcommand{\printifnonempty}[2]{\if\relax\detokenize{#1}\relax\else#2\fi}
\newcommand{\alter}[5]{%
  \long\def\temp{#3}%
  \long\def\accept{#5}%
  \ifx\temp\accept
    {#1}
  \else
    {\textcolor{#4}{\printifnonempty{#1}{{#1}}}%
    \textcolor{grey}{\printifnonempty{#2}{(#2)}}%
    \textcolor{#4}{\printifnonempty{#3}{{[#3]}}}}%
  \fi
}

\newcommand{\beq}{\begin {equation}}  
\newcommand{\eeq}{\end   {equation}} 
\newcommand{\bea}{\begin {eqnarray}} 
\newcommand{\eea}{\end   {eqnarray}}  
\newcommand{\baa}{\begin {array}   } 
\newcommand{\eaa}{\end   {array}   }     
\newcommand{\bit}{\begin {itemize} }
\newcommand{\eit}{\end   {itemize} }
\newcommand{\be }{\begin {equation}} 
\newcommand{\ee }{\end   {equation}}

\begin{document}
\preprint{FERMILAB-PUB-18-381-T}
\preprint{OSU-HEP-18-05}

\title{Neutrino Masses and Mixings Dynamically Generated by a Light Dark Sector}

\author{Enrico Bertuzzo}
\email[E-mail:]{bertuzzo@if.usp.br}
 \affiliation{Departamento de F\'isica Matem\'atica, Instituto de F\'isica\\
Universidade de S\~ao Paulo, C.P. 66.318, S\~ao Paulo, 05315-970, Brazil}
\author{Sudip Jana}%
 \email[E-mail:]{sudip.jana@okstate.edu}
\affiliation{Department of Physics and Oklahoma Center for High Enrgy Physics,\\
 Oklahoma State University, Stillwater, OK 74078-3072, USA}%
\affiliation{Theory Department, Fermi National Accelerator Laboratory, P.O. Box 500,
Batavia, IL 60510, USA}%
\author{Pedro A. N. Machado}
\email[E-mail:]{pmachado@fnal.gov}
\affiliation{Theory Department, Fermi National Accelerator Laboratory, P.O. Box 500,
Batavia, IL 60510, USA}%
\author{Renata Zukanovich Funchal}
\email[E-mail:]{zukanov@if.usp.br}
 \affiliation{Departamento de F\'isica Matem\'atica, Instituto de F\'isica\\
Universidade de S\~ao Paulo, C.P. 66.318, S\~ao Paulo, 05315-970, Brazil}

\begin{abstract}
\vspace{0.3 in}
\noindent
Neutrinos may be the harbingers of new dark sectors, since the renormalizable neutrino portal allows for their interactions with hidden new physics. We propose here to use this fact to connect the generation of neutrino masses to a light dark sector, charged under a new $U(1)_{\cal D}$ dark gauge symmetry. We introduce the minimal number of dark fields to obtain an anomaly free theory with spontaneous breaking of the dark symmetry, and obtain automatically the inverse seesaw Lagrangian. In addition, the so-called $\mu$-term of the inverse seesaw is dynamically generated and technically natural in this framework. As a bonus, the new light dark gauge boson can provide a possible explanation to the MiniBooNE anomaly.  
\end{abstract}

\maketitle

\section{\label{sec:Introduction}Introduction}

One of the most surprising experimental results of the last decades has been the 
discovery of tiny neutrino masses and relatively large neutrino mixings. 
Although non-vanishing neutrino masses are a clear indication of physics beyond 
the Standard Model (SM), the mechanism and the scales responsible for the neutrino mass generation remain a total mystery.

It seems unlikely that the very small neutrino masses are generated by the 
same Higgs mechanism responsible for the masses of the other SM fermions, since extremely small Yukawa couplings, of the order of $\lesssim 10^{-12}$, must be invoked. 
A more `natural' way to generate neutrino masses involve the addition of new states that, once integrated out, generate the dimension five Weinberg operator
\begin{equation}
  \mathcal{O}_5=\frac{c}{\Lambda}LLHH.
\end{equation}
This is embodied by  the so-called seesaw mechanisms~\cite{seesawI,seesawII,seesawIII, Mohapatra:1986bd}.
The smallness of neutrino masses relative to the weak scale implies either that the scale of new physics $\Lambda$ is very large (making it impossible to experimentally discriminate the different seesaw mechanisms), or that the Wilson coefficient $c$ is extremely small (for instance, coming from loop effects involving singly or doubly charged scalars~\cite{radiativeseesaw}).

A different approach is given by neutrinophilic Two-Higgs-Doublet Models~\cite{Gabriel:2006ns, Davidson:2009ha}. In this framework, a symmetry ($U(1)$ or $Z_2$) compels one of the doublets to couple to 
all  SM fermions but neutrinos, hence being responsible for their masses, while the other Higgs 
couples to the lepton doublets and right-handed neutrinos. If the second doublet
acquires a vacuum expectation value (vev) around the eV scale, this leads to 
small neutrino masses. These models, however, are either ruled out or severely constrained by electroweak precision data and low energy flavor physics~\cite{Machado:2015sha,Bertuzzo:2015ada}.

A variation of this idea, in which the symmetry is taken 
to be a local $U(1)$ and leads to the typical Lagrangian of the inverse seesaw 
scenario,  suffers from an accidental lepton number symmetry that has to be 
explicitly broken to avoid the presence of a massless Nambu-Goldstone boson in the spectrum~\cite{Bertuzzo:2017sbj}.
All aforementioned models have one of the two following features: (i) The model is realized at  very high scales, or (ii) the model is based on explicit breaking of lepton number or other symmetries that protect neutrino masses (e.g. in TeV scale type II or inverse seesaw models). 

Neutrinos, however, are the {\em darkest} between the SM particles, in
the sense that they can couple through the renormalizable {\em
  neutrino portal} $LH$ operator with generic dark
sectors~\cite{Schmaltz:2017oov}. This fact has been used in connection
to thermal Dark Matter with mass in the sub-GeV region (see for
instance Refs.~\cite{Falkowski:2011xh,Batell:2017cmf}). In this letter
we propose to use such a portal to explicitly connect a new light dark
sector with the generation of neutrino masses. In this way, we are
able to lower the scale of neutrino mass generation below the
electroweak one by resorting to a dynamical gauge symmetry breaking of
this new sector. The dark sector is mostly secluded from experimental
scrutiny, as it only communicates with the SM by mixing among scalars,
among neutrinos and dark fermions, and through kinetic and mass mixing
between the gauge bosons.  This scheme has several phenomenological
consequences at lower energies, and in particular it offers a natural
explanation for the long-standing excess of electron-like events
reported by the MiniBooNE collaboration~\cite{Aguilar-Arevalo:2018gpe,
  Bertuzzo:2018itn}.

\section{\label{sec:model} The Model}

To avoid any neutrino mass contribution from the Higgs mechanism, we introduce 
a new dark gauge symmetry $U(1)_{\cal D}$, under which the SM particles are uncharged, 
but the new sector is charged.  
To build a Dirac neutrino mass term we need a $SU(2)_L$ singlet right-handed dark neutrino $N$, and  a dark scalar doublet $\phi$, both having the same $U(1)_{\cal D}$ charge $+1$. 
The absence of chiral anomalies require a second 
right-handed neutrino, $N'$, with an opposite $U(1)_{\cal D}$ charge, thus restoring lepton number symmetry. 
We add to the particle content a dark  scalar $SU(2)_L$ singlet $S_{2}$, with dark charge $+2$, whose vev spontaneously breaks lepton number, giving rise to a Majorana mass component for the dark neutrinos. As we will see shortly, this setup leads to an inverse seesaw structure in which the lepton number breaking parameter is promoted to a dynamical quantity. Finally, this scalar content enjoys an accidental global symmetry which is spontaneously broken. To avoid a massless Goldstone boson, an extra dark scalar $SU(2)_L$ singlet $S_{1}$, with dark charge $+1$, is included in the spectrum. Its vev breaks all accidental global symmetries. 
This field will allow for mixing among all the scalar fields, including the SM Higgs. 

The dark scalar $S_{1}$ will spontaneously break $U(1)_{\cal D}$ by acquiring a vev, while $\phi$ and $S_2$ 
will only develop an induced vev after the breaking of the electroweak and  dark symmetries. 
By making a well motivated choice for the hierarchy of the vevs, our model allows a dynamical generation of the light neutrino masses and mixings at very low scale.
Our model predicts masses for the dark scalars within the reach of current experiments as well as a light dark vector boson, $Z_{\cal D}$, that has small kinetic mixing with the photon and mass mixing with the SM $Z$ boson.

The dark particles communicate with the SM ones via mixing: flavor 
mixing (neutrinos), mass mixing (scalars) and  mass mixing and kinetic 
mixing ($Z_{\cal D}$), giving rise to a simple yet rich phenomenology.
\vspace{0.3cm}

\subsection{\label{subsec:scalars} The Dark Scalar Sector}
Let us start discussing the scalar sector of the model. This will motivate the region of parameter space on which we will focus throughout the paper. The most general $SU(2)_L\times U(1)_{Y} \times U(1)_{\cal D}$
invariant scalar potential that can be constructed out of the fields and charges outlined above is \begin{widetext}
\begin{align}\begin{aligned}\label{eq:potential}
  V = &-m_H^2(H^\dagger H)+m^2_{\phi}(\phi^\dagger\phi)
  	-m_{1}^2S_{1}^*S_{1} + m_{2}^2 S_{2}^* S_{2} -\left[\frac{\mu}{2} S_{1}(\phi^\dagger H)+\frac{\mu'}{2}S_{1}^2 S_{2}^* + \frac{\alpha}{2}(H^\dagger \phi)S_{1} S_{2}^*+{\rm h.c.} \right]  \\
	& ~~~~~+ \lambda_{H\phi}' \phi^\dagger H H^\dagger \phi + \sum_\varphi^{\{H,\phi,S_1,S_{2}\}}\lambda_\varphi(\varphi^\dagger \varphi)^2 + \sum_{\varphi<\varphi'}^{\{H,\phi,S_1,S_2\}}\lambda_{\varphi\varphi'}(\varphi^\dagger \varphi)(\varphi^{\prime\dagger} \varphi')\, .
\end{aligned}\end{align}
\end{widetext}
(In the last sum, the notation $\varphi<\varphi'$ is to avoid double counting.)
We denote the vevs of the scalar fields as 
$(H , \phi  , S_1 ,  S_2)|_{\rm vev}\equiv\left(v,v_\phi,\omega_1,\omega_2\right)/\sqrt{2}$.
We stress that we are supposing the bare mass terms of $H$ and $S_1$ to be negative, while we take the corresponding ones for $\phi$ and $S_2$ to be positive. This ensures that, as long as $\mu=\mu'=\alpha \equiv 0$ ({\it i.e.} if there is no mixing among the scalar fields), the latter fields do not develop a vev, while the former do. In turn, this implies that the vevs $v_\phi$ and $\omega_2$ must be induced by $\mu$, $\mu'$, and/or $\alpha$. 

We now observe that $\mu$, $\mu'$, and $\alpha$ explicitly break two
accidental $U(1)$ global symmetries, making these parameters
technically natural~\footnote{ One of the symmetries is lepton number,
  the other is a symmetry under which only $\phi$ and $L$ are charged,
  with opposite charge. Since there are only two global symmetries for
  3 parameters, having two of them non-zero necessarily generates the
  third by renormalization group running. }. For our purposes, this
means that $\mu$, $\mu'$ and $\alpha$ can be taken small in a natural
way, and justifies our working hypothesis $v_\phi,\omega_2\ll
v,\omega_1$.  As we will see later, this hierarchy of vevs will
provide a low scale realization of the inverse seesaw mechanism with
low scale dynamics associated to it.  Explicitly, we obtain
\begin{align}\label{eq:vevs}
  &v_\phi\simeq \frac{1}{8\sqrt{2}}\left(\frac{\alpha\mu' \, v \omega_1^3}{M_{S'_{\cal D}}^2M_{H_{\cal D}}^2}+4\frac{\mu \, \omega_1 v}{M_{H_{\cal D}}^2}\right)\, ,~~~~~{\rm and}\\ %
  &\omega_2\simeq \frac{1}{8\sqrt{2}}\left(\frac{\alpha\mu \, v^2\omega_1^2}{M_{S'_{\cal D}}^2 M_{H_{\cal D}}^2}+4\frac{\mu' \, \omega_1^2}{M_{S'_{\cal D}}^2}\right) \, ,
\end{align}
with $M_{H_{\cal D}}^2$ and $M_{S'_{\cal D}}^2$ approximately being the physical masses of the respective scalars (to be defined below). 
In order to avoid large mixing between $H$ and $\phi$, we will always make the choice $\omega_1 \ll v$.

The scalar spectrum contains, in addition to the SM-like scalar $h_{\rm SM}$ with mass  $m_{h_{\rm SM}}\simeq $ 125 GeV, three CP-even dark scalars $H_{\cal D}$, $S_{\cal D}$ and 
$S'_{\cal D}$, with masses $M_{H_{\cal D}}$, $M_{S_{\cal D}}$ and 
$M_{S'_{\cal D}}$, two CP-odd dark scalars $A_{\cal D}$ and $a_{\cal D}$ 
with masses $M_{A_{\cal D}}$ and $M_{a_{\cal D}}$, and a charged dark scalar $H^\pm_{\cal D}$ with mass $M_{H^\pm_{\cal D}}$.

Explicitly, the masses of the CP-even scalars are~\footnote{Radiative
  corrections will naturally contribute to the masses of these
  scalars.  There are potentially several contributions according to
  Eq.~(\ref{eq:potential}), the quartic couplings being the most
  dangerous ones. In order to avoid fine-tuning{, we will always
    demand the masses of the lightest scalars to satisfy $M_{\rm lightest}
    \gtrsim \sqrt{\lambda} M_{\rm heavy}/8\pi$, where $M_{\rm heavy}$ denotes
    any of the heavy scalar masses.} By the same argument we expect $\mu, \mu'$
  and $\alpha v$ to be below $M_{\rm lightest}$. { Our computation
    ignores the threshold at the Planck scale, which must be
    stabilized by other means (for instance, supersymmetrizing the
    theory).}}
\begin{align}\begin{aligned}\label{eq:scalar_masses}
  m_{h_{\rm SM}}^2 &\simeq 2 \, \lambda_H v^2\, , \\
  M_{S_{\cal D}}^2 &\simeq 2 \, \lambda_{S_{1}} \omega_1^2\, , \\
  M_{H_{\cal D}}^2 &\simeq m_{\phi}^2 + \frac{\lambda_{H\phi}+\lambda_{H\phi}'}{2} v^2 \, ,\\
  M_{S'_{\cal D}}^2 & \simeq m_{2}^2 + \frac{\lambda_{H S_2}}{2} \, v^2 \, ,\\
\end{aligned}\end{align}
while the masses of the CP-odd and charged scalars are given by
\begin{align}
M_{A_{\cal D}} & \simeq M_{H_{\cal D}}\, , \\
M_{a_{\cal D}} & \simeq M_{S'_{\cal D}}\, , \\
M_{H_{\cal D}^\pm}^2 & \simeq M_{H_{\cal D}}^2 - \frac{\lambda_{H\phi }' v^2}{2}\, .
\end{align}

As for the composition of the physical states, since the mixing in the scalar sector is typically small, we can generically define
\begin{equation}
  \varphi_{\rm physical} = \varphi - \sum_{\varphi'\neq \varphi}\theta_{\varphi \varphi'} \varphi'\,,
\end{equation}
where $\varphi_{\rm physical}$ denotes the physical scalar field that has the largest $\varphi$ admixture. 
Then, the mixing in the CP-even scalar sector is given by 
\begin{align}\label{eq:scalar_mixing}
  \theta_{H \phi } &\simeq  \, \left[(\lambda_{H\phi}+\lambda_{H\phi }')\, v_\phi v - \mu\omega_1/2\sqrt{2}\right]/\Delta M^2_{h_{SM}H_{\cal D}}\, , \nonumber \\
 \theta_{H {S_1} } &\simeq  \, \lambda_{H S_1}\, \omega_1 v/\Delta M^2_{h_{SM}S_{\cal D}}\, , \nonumber \\
  \theta_{H {S_2} } &\simeq  \lambda_{HS_2}\, \omega_2 v/\Delta M^2_{h_{SM} S'_{\cal D}} \, , \nonumber\\
   \theta_{\phi {S_1} } &\simeq \mu v /2\sqrt{2}\Delta M^2_{H_{\cal D}S_{\cal D}}  \, ,\\
    \theta_{\phi {S_2} } &\simeq  \alpha \, \omega_1 v/4\Delta M^2_{H_{\cal D}S'_{\cal D}}  \, \nonumber ,\\
  \theta_{{S_1} {S_2} } & \simeq  \mu'\omega_1/\sqrt{2}\Delta M^2_{S_{\cal D}S'_{\cal D}}  \, , \nonumber 
\end{align}
where $\Delta M^2_{\varphi \varphi'}\equiv M^2_\varphi-M^2_{\varphi'}$,
while the  Nambu-Goldstone bosons associated with the $W^\pm$, $Z$ and $Z_{\cal D}$ bosons are defined as
\begin{align}
 G_W^\pm &\simeq H^\pm - \frac{v_\phi}{v}\phi^\pm \,, \nonumber\\
 G_Z & \simeq {\rm Im}(H^0) + \frac{v_\phi}{v} {\rm Im}(\phi^0)\,,\\
 G_{Z_{\cal D}}&\simeq{\rm Im}(S_1)+\frac{2\omega_2}{\omega_1}{\rm Im}(S_2)+\frac{v_\phi}{\omega_1}{\rm Im}(\phi^0)-\frac{v_\phi^2}{\omega_1 v}{\rm Im}(H^0).\nonumber
\end{align}

We see that our hypothesis $v_\phi, \omega_2 \ll \omega_1 \ll v$ prevents any relevant modification to the Higgs-like couplings, and $h_{\rm SM}$ ends up being basically like the SM Higgs boson.
Moreover, due to the mixing with the Higgs field, the dark scalars and the longitudinal mode of the $Z_{\cal D}$ will also couple to SM fermions via  SM Yukawa couplings. Nevertheless, such couplings to light fermions are quite small as they are suppressed by the hierarchy of vevs. If the spectrum enjoys light degrees of freedom (below the $100$ MeV scale), an interesting phenomenology may be associated with this sector. A dedicated study will be pursued in a future work.

\subsection{\label{subsec:neutrinos} Neutrino Masses and Mixings}
Let us now discuss the generation of neutrino masses and mixings, and how the dynamics of the dark sector outlined so far ensures light neutrinos. The most general Lagrangian in the neutrino sector, compatible with our charge assignment, is
\begin{align}
  \mathcal{L}_\nu=& -y_\nu \, \overline{L} \widetilde{\phi} N + y_N \, S_2 \,\overline{N}N^c + y_{N'}\, S_{2}^* \, \overline{N'}N'^c \nonumber\\
  &+ m\, \overline{N'}N^c+{\rm h.c.}\, ,
  \label{eq:Lnu}
\end{align}
where $y_\nu$, $m$, $y_N$ and $y_{N'}$ are matrices in flavor space. After the two-steps spontaneous breaking $SU(2)_L \times U(1)_Y \times U(1)_{\cal D}\xrightarrow[\quad ]{v} U(1)_{\rm em} \times U(1)_{\cal D} \xrightarrow[\quad ]{\omega_1} U(1)_{\rm em}$, the neutrino mass matrix 
in the $(\nu,\,N,\,N')$ basis is
\begin{equation}
  \mathcal{M}_\nu=\frac{1}{\sqrt{2}}\left(
  	\begin{array}{c c c}
	0			&  y_\nu \, v_\phi  	&  0\\
	y_\nu^T \, v_\phi  &  y_N \, \omega_2	&  \sqrt{2}\,m\\
	0 			&  \sqrt{2}\,m^T				&y_{N'} \,\omega_2
	\end{array}\right)\, .
\end{equation}
As already stressed, $v_\phi$ generates a Dirac mass term, while $\omega_2$ plays the key role to generate a naturally small term $y_{N^{\prime}}\omega_2$, which can be identified as the tiny mass term of the inverse seesaw $\mu_{\rm ISS}$ (the dimensionful parameter of inverse seesaw that breaks lepton number by two units), and we obtain a dynamically generated inverse seesaw neutrino mass matrix. The mass matrix $m$ can always be made diagonal, and in principle take any value, but given the smallness of the Dirac and $\mu_{\rm ISS} $-terms, it is clear that we can generate light neutrino masses even with values of $m$ smaller than that in the usual inverse seesaw scenario. 

More precisely, the light neutrino mass matrix is given at leading order by
\begin{equation}
  m_\nu \simeq (y_\nu^T v_\phi) \frac{1}{m^T} (y_{N^{\prime}} \omega_2) \frac{1}{m} (y_\nu v_\phi)\, .
\end{equation}
\begin{figure}[t]
\includegraphics[width=0.5\textwidth]{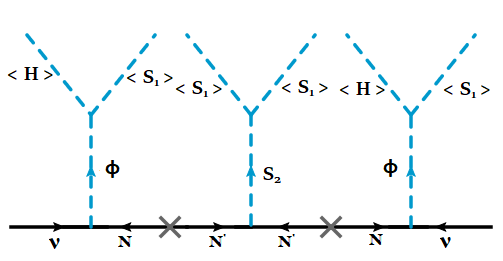}
\caption{\label{fig:massdiagram}
Diagram for the dynamically induced light neutrino masses in our model.}
\end{figure}
Inspection of Eq.~(\ref{eq:Lnu}) makes clear why we can substantially lower the 
scale of neutrino mass generation, since in our construction the light neutrino masses are generated effectively as a  dimension nine operator (see Fig.~\ref{fig:massdiagram}). Schematically, we start with
\begin{equation}
\label{eq:dsix}
{\cal L}_\nu^{\rm eff} \sim y_\nu^2 \, y_{N^{\prime}}\frac{(\overline{L^c} \phi)(\phi^T L)}{m^2} S_{2}^* \, .
\end{equation}
Remembering that the vevs of $\phi$ and $S_2$ are induced by the dynamics of the scalar sector, we can rewrite the previous operator in terms of $H$ and $S_1$, the fields whose vev's are present even in the limit $\left\{ \mu, \mu', \alpha\right\}\to 0$. We obtain
\begin{equation}
\label{eq:dnine}
{\cal L}_\nu^{\rm d=9} \sim y_\nu^2 \, y_{N^{\prime}} \frac{\mu^2}{M^4_{H_{\cal D}}} \frac{\mu'}{M^2_{S'_{\cal D}}}\frac{(\overline{L^c} H)(H^T L)}{m^2} (S_{1}^* S_{1})^2 \, ,
\end{equation}
from which it is clear that, ultimately, neutrinos masses are generated by a dimension 9 operator (see, e.g., Refs.~\cite{higherdim} for generation of neutrino masses from higher dimensional effective operators). In addition, we have a further suppression due to the fact that $\mu$ and $\mu'$ can be taken small in a technically natural way.

The mixing between active and dark neutrinos can be explicitly written as 
\begin{equation}\label{eq:mix}
\nu_\alpha = \sum_{i=1}^{3} U_{\alpha i} \, \nu_i + U_{\alpha \mathcal{D}} \, N_{\cal D}\, ,
\end{equation}
$\alpha=e,\mu,\tau,{\cal D}$, where $\nu_i$ and $\nu_{\alpha}$ are the neutrinos mass and flavor eigenstates, respectively (we denote by $\alpha = {\cal D}$ the 6 dark neutrinos flavor states, while $U_{\alpha \mathcal{D}}$ is a $9\times 6$ matrix). Schematically, we have that the mixing between light and heavy neutrinos is $y_\nu v_\phi/m$. Note that the dark neutrino can be made very light, without introducing too large mixing, even for $y_\nu\sim\mathcal{O}(1)$ since $v_\phi\ll v$.
\vspace{0.3cm}

\subsection{\label{subsec:gauge} $Z_{\cal D}$ and the Gauge Sector}
The new vector boson will, in general, communicate with the SM  sector via
either mass mixing or kinetic mixing. The relevant part of the dark Lagrangian is
\begin{widetext}
\begin{equation}
{\cal L}_{\cal D} \supset \frac{m^2_{Z_{\cal D }}}{2} \,
Z_{{\cal D}\mu} Z_{\cal D}^{\mu} + g_{\cal D} Z_{\cal D}^\mu \, J_{\mathcal{D}\mu} + e \epsilon \, Z_{\cal D}^\mu \,
J_\mu^{\rm em}
+ \frac{g}{c_W} \epsilon' \, Z_{\cal D}^\mu \,
J_\mu^{\rm Z} \, ,
\label{eq:kmix}
\end{equation}
\end{widetext}
 where $m_{Z_{\cal D}}$ is the mass of $Z_{\cal D}$, $g_{\cal D}$ is the $U(1)_{\cal D}$ gauge coupling, $e$ is the electromagnetic coupling, $g/c_W$ is the $Z$ coupling in the SM, while $\epsilon$ and $\epsilon'$ parametrize the kinetic and mass mixings, respectively. The electromagnetic and $Z$ currents are denoted by $\
J^{\rm em}_\mu$ and $J^{Z}_\mu$, while $J_{\mathcal{D}\mu}$ denotes the dark current.

In the limits we are considering, the $Z$ and $W^\pm$ masses are essentially unchanged with respect to the SM values, while the new gauge boson mass reads
\begin{equation}
  m_{Z_{\cal D}}^2\simeq g_{\cal D}^2 \left(\omega_1^2 +  v_\phi^2 + 4 \, \omega_2^2\right)\simeq g_{\cal D}^2 \, \omega_1^2\, ,
\end{equation}
with mass mixing between $Z$ and $Z_{\cal D}$ given by
\begin{align}
	\epsilon' \simeq \frac{2 g_{\cal D}  }{g/c_W} \frac{v_\phi^2}{v^2}\, .
\end{align}
Of course, a non-vanishing mass mixing $\epsilon'$ implies that the $Z$ boson inherits a coupling to the dark current
\begin{align}
  {\cal L}_{Z} = \frac{m_Z^2}{2} Z_\mu Z^\mu + \frac{g}{c_W} Z^\mu J_\mu^Z - g_{\cal D} \epsilon' Z^\mu J_{\mathcal{D}\mu}\, .
\end{align}
While the new coupling allows for the possibility of new invisible $Z$ decays, the large hierarchy $v_\phi \ll v$ guarantees that the new contributions to the invisible decay width are well inside the experimentally allowed region. The vev hierarchy also protects the model from dangerous $K$, $B$ and $\Upsilon$ decays with an on-shell $Z_{\cal D}$ in the final state~\cite{Davoudiasl:2012ag, Babu:2017olk}.

The kinetic mixing parameter $\epsilon$ is allowed { at tree-level} by all symmetries of the model. Moreover, it is radiatively generated (see e.g. Ref.~\cite{Holdom:1985ag})  by a loop of the $H^\pm_{\cal D}$ scalar  which magnitude is
\begin{align}
\epsilon_{\rm LOOP} \sim \frac{e g_{\cal D}}{480 \pi^2} \frac{m_{Z_{\cal D}}^2}{m_{H^\pm_{\cal D}}^2}.
\end{align}
{ In what follows, we will take $\epsilon$ as generated at tree-level, with $\epsilon_{\rm TREE} \gg \epsilon_{\rm LOOP}$ to guarantee the radiative stability of the theory.} 
 The kinetic mixing will lead to interactions of the $Z_{\cal D}$ to charged fermions, as well as decays if kinematically allowed (see e.g. Ref.~\cite{Ilten:2018crw} for constraints).

\section{\label{sec:pheno}Phenomenological Consequences}
We would like at this point to make some comments about the possible 
phenomenological consequences of our model. To illustrate the discussion
let us consider a benchmark point consistent with our working 
hypothesis $v_{\phi}, \omega_2 \ll \omega_1 \ll v$. This point is defined by the 
input values given in Tab.~\ref{tab1}, producing the physical observables 
in Tab.~\ref{tab2}.
 
 We see that for this point the changes in the masses of the SM gauge
 bosons as well as the mixings of the Higgs with the new scalars are
 negligible, so we do not foresee any major problems to pass the
 constraints imposed to the SM observables by the Tevatron, LEP or the
 LHC data.  { Moreover, our model is endowed with all the
   features needed to explain the excess of electron-like events
   observed by the MiniBooNE experiment: a new dark vector boson,
   $Z_{\cal D}$, that couples to the SM fermions by kinetic mixing and
   also directly to a dark neutrino, $\nu_{\cal D}$, which mixes with
   the active ones.  As shown in \cite{Bertuzzo:2018itn}, the dark
   neutrino can be produced via neutrino-nucleus scattering in the
   MiniBooNE detector and, if $m_{N_{\cal D}}>m_{Z_{\cal D}}$,
   subsequently decay as $N_{\cal D} \to Z_{\cal D} + \nu_i$. The
   $Z_{\cal D}$ can then be made to decay primarily to $e^+ e^-$ pairs
   with a rate that results in an excellent fit to MiniBooNE energy
   spectra and angular distributions.}

In general, this model may in principle also give contributions to the muon $g-2$ \footnote{Since  additional  electrically charged/neutral scalar ($H^\pm_{\cal D}, H_{\cal D}, A_{\cal D}$)  fields and a light dark gauge boson($Z_{\cal D}$) field  are  present  in  our  model,  they will induce a shift in the leptonic magnetic moments and mediate LFV decays via the interactions as shown in Eq.~\ref{eq:Lnu} and Eq.~\ref{eq:kmix}. The contribution to muon magnetic moment from neutral dark Higgs fields ($H_{\cal D}, A_{\cal D}$) with flavor-changing couplings is negligible in our framework. The dominant contribution will arise from singly charged scalar ($H^\pm_{\cal D}$) via the interaction term $y_\nu \, \overline{L} \widetilde{\phi} N$. But, the singly charged scalar correction to muon $g-2$ is negative and rather destructive to the other contributions. Whereas, the one loop contribution of the dark gauge boson ($Z_{\cal D}$) to muon $g-2$ is quite promising and a dedicated study will be pursued further on that. It is worth mentioning that there will be another small contribution to muon $g-2$ from the $W$ boson exchange via mixing between active and dark neutrinos.}, to atomic parity violation, polarized electron scattering, neutrinoless double $\beta$   decay, rare meson decays as well as  to other low energy observables such as  
 the running of the weak mixing angle $\sin^2\theta_{W}$. 
 There might be consequences to neutrino experiments too. It can, for instance, 
 modify neutrino scattering, such as coherent neutrino-nucleus scattering, or 
 impact neutrino oscillations experimental results
 as this model may give rise to  non-standard neutrino interactions in matter.
 Furthermore, data from  accelerator neutrino experiments, such as MINOS, NO$\nu$A, T2K, and MINER$\nu$A, may be used to probe $Z_{\cal D}$ decays to charged leptons, in particular, if the channel $\mu^+\mu^-$ is kinematically allowed. 
 We  anticipate new rare Higgs decays, such as $h_{\rm SM} \to Z Z_{\cal D}$, or 
 $H^\pm_{\cal D} \to W^\pm Z_{\cal D}$, that depending on $m_{Z_{\cal D}}$ may affect LHC  
physics. Finally, it may be  interesting to examine the apparent anomaly seen in $^8$Be decays~\cite{Kozaczuk:2016nma} in the light of this new dark sector.

The investigation of these effects is  currently under way but  beyond the scope of this letter and shall be presented  in a future work.
 
\begin{table}[htb]
\begin{tabular}{||c|c|c|c||}
\hline\hline
\multicolumn{4}{c}{\bf Vacuum Expectation Values}\\
\hline \hline
$v$ (GeV) & $\omega_1$ (MeV) & $v_{\phi}$ (MeV) & $\omega_2$ (MeV)\\
\hline
$246$ & $136$ &   $0.176$ & $0.65$ \\ 
\hline \hline
\multicolumn{4}{c}{\bf Coupling Constants}\\
\hline\hline
$\lambda_H$ & $\lambda_{H\phi}=\lambda_{H\phi}'$ & $\lambda_{H S_1}$ & $\lambda_{H S_2}$ \\
\hline
$0.129$ & $10^{-3}$ & $10^{-3}$ & $-10^{-3}$  \\
\hline
$\lambda_{\phi S_1}$ & $\lambda_{\phi S_2}$ & $\lambda_{S_1}$ & $\lambda_{S_1 S_2}$ \\
\hline
$10^{-2}$ & $10^{-2}$ & $2$ & $0.01$  \\
\hline
$\mu$ (GeV) & $\mu'$ (GeV) & $\alpha$ & $g_{\cal D}$\\
\hline
$0.15$ & $0.01$ &$10^{-3}$ & 0.22 \\
\hline\hline
\multicolumn{4}{c}{\bf Bare Masses}\\
\hline\hline
\multicolumn{2}{||c|}{$m_{\phi}$ (GeV)} & \multicolumn{2}{c||}{$m_2$ (GeV)}  \\
\hline \hline
\multicolumn{2}{||c|}{100} & \multicolumn{2}{c||}{5.51}  \\
\hline \hline
\end{tabular}
\caption{\label{tab1} Input values for a benchmark point in our model that can provide an explanation of the low energy MiniBooNE excess~\cite{Aguilar-Arevalo:2018gpe,Bertuzzo:2018itn}. See Tab.~\ref{tab2} for the respective physical masses and mixings.}
\end{table}

\begin{table*}[htb]
\begin{tabular}{||c|c|c|c|c|c|c|c|c||}
\hline\hline
\multicolumn{9}{c}{\bf Masses of the Physical Fields}\\
 \hline\hline
$m_{h_{\rm SM}}$ (GeV) & $m_{H_{\cal D}}$ (GeV) & $m_{S_{\cal D}}$ (MeV) &  $m_{S'_{\cal D}}$ (MeV) &  $m_{H^\pm_{\cal D}}$ (GeV) & $m_{A_{\cal D}}$ (GeV)  & $ m_{a_{\cal D}}$ (MeV) & $m_{Z_{\cal D}}$ (MeV) & $m_{N_{\cal D}}$ (MeV)\\
\hline
$125$ & $100$ &   $272$ & $320$ & 100 & 100 & 272 & 30  & 150 \\ 
\hline \hline
\multicolumn{9}{c}{\bf Mixing between the Fields }\\
 \hline\hline
 $\theta_{H\phi} $  & $\theta_{HS_1}$ & $\theta_{HS_2}$ &  $\theta_{\phi S_1}$ &  $\theta_{\phi S_2}$ & $\theta_{S_1 S_2} $  &  $e\epsilon$ & $\epsilon' $ & $|U_{\alpha N}|^2$\\
\hline
$1.3\times 10^{-6}$ & $2.1 \times 10^{-6}$ &   $10^{-8}$ & $1.2 \times 10^{-3}$ & $8.3 \times 10^{-7}$ & $3.4\times 10^{-2}$ & $2\times 10^{-4}$ & $3.6 \times 10^{-14}$ & $
\mathcal{O}(10^{-6})$  \\ 
\hline \hline
\end{tabular}
\caption{\label{tab2} Physical masses and mixings for the benchmark point of our model that can provide an explanation of the low energy MiniBooNE excess~\cite{Aguilar-Arevalo:2018gpe,Bertuzzo:2018itn}. The light-heavy neutrino mixing is schematically denoted by $|U_{\alpha N}|^2$, and $m_{N_{\cal D}}$ denotes the order of magnitude of the diagonal entries of the dark neutrino mass matrix.}
\end{table*}

\section{\label{sec:Conclusion}Final Conclusions and Remarks}

The main purpose of this letter has been to explicitly connect the generation of neutrino masses to the existence of a new light dark sector. Doing so, we are able to lower the scale of neutrino mass generation substantially below the electroweak one by resorting to a dynamical breaking of a new $U(1)_{\cal D}$ dark gauge symmetry under which SM particles are neutral.

Our secluded sector consists of the minimal dark field content able to ensure anomaly cancellation, as well as the spontaneous breaking of the dark gauge symmetry without the appearance of a Nambu-Goldstone boson. It consists of a new scalar doublet, two scalar singlets and a set of six new fermion singlets, all charged under the dark symmetry. A judicious choice of dark charges allows to generate neutrino masses by a dynamical inverse seesaw mechanism, but unlike  the usual inverse seesaw scenario, the so-called $\mu_{\rm ISS}$-term is here dynamically generated, and can be small in a technically natural way. Interestingly, neutrino masses effectively emerge only at the level of dimension 9 operators, and we can have a new light dark gauge boson in the spectrum.

The dark sector is mostly secluded from experimental scrutiny, as it only communicates with the SM by mixing: the SM Higgs mixing  with dark scalars,  neutrinos 
mixing with dark fermions, and through kinetic and mass mixing with the dark gauge boson.

The low scale phenomenology of the model is simple yet rich. It is possible that 
our model gives sizable contributions to  several experimental observables 
such as the value of the muon $g-2$,  the Majorana mass in neutrinoless double 
$\beta$  decay or influence atomic parity violation, polarized electron scattering, 
or rare meson decays, among others.
Moreover, the mechanism we propose in this letter could provide an novel explanation to the MiniBooNE low energy excess of electron-like events~\cite{Bertuzzo:2018itn}.

As a final remark, let us stress that we presented here only the low scale realization of the model, 
imposed by the hierarchy of vevs we have selected.  Nevertheless, we could have chosen a different one, for instance, $\omega_1\gtrsim v$. In that case we would have a high scale realization of the model, with unique phenomenological consequences at the LHC, for instance displaced vertex or prompt multi-lepton signatures.

\section*{Acknowledgments}
\noindent
We thank Kaladi Babu for useful discussions and Oscar \'{E}boli for careful reading of the manuscript. This work was supported by Funda\c{c}\~ao de Amparo \`a Pesquisa do Estado de S\~ao Paulo (FAPESP) under contracts 12/10995-7 and 2015/25884-4, and by Conselho Nacional de Ci\^encia e Tecnologia (CNPq). 
R.Z.F. is grateful for the hospitality of the Fermilab Theoretical Physics Department
during the completion of this work.
The work of S.J. is supported in part by the US Department of Energy Grant 
(DE-SC0016013) and the Fermilab Distinguished Scholars Program. Fermilab is 
operated by the Fermi Research Alliance, LLC under contract No. DE-AC02-07CH11359 with the United States Department of Energy. This project has also received support from the European Union's Horizon 2020 research and innovation programme under the Marie Sklodowska-Curie grant agreement N$^\circ$ 690575 (InvisiblesPlus) and 
N$^\circ$ 674896 (Elusives).                                      



\end{document}